\documentclass[aps,pra,twocolumn,showpacs,superscriptaddress,letterpaper,longbibliography]{revtex4-1}

\usepackage{graphicx}
\usepackage{mdframed}
\usepackage{framed}
\usepackage{indentfirst}
\usepackage{natbib}
\usepackage{amsmath,amssymb}
\usepackage{subfigure}
\usepackage{enumitem}
\usepackage{epstopdf}
\usepackage{xfrac}
\usepackage{multirow}

\begin{document}

\title{Improvement or selection?  A longitudinal analysis of students' views about experimental physics in their lab courses}


\author{Bethany R. Wilcox}
\affiliation{Department of Physics, Colorado School of Mines, Golden, CO 80401}

\author{H. J. Lewandowski}
\affiliation{Department of Physics, University of Colorado, 390 UCB, Boulder, CO 80309}
\affiliation{JILA, National Institute of Standards and Technology and University of Colorado, Boulder, CO 80309}

\begin{abstract}
Laboratory courses represent a unique and potentially important component of the undergraduate physics curriculum, which can be designed to allow students to authentically engage with the process of experimental physics.  Among other possible benefits, participation in these courses throughout the undergraduate physics curriculum presents an opportunity to develop students' understanding of the nature and importance of experimental physics within the discipline as a whole.  Here, we present and compare both a longitudinal and pseudo-longitudinal analysis of students' responses to a research-based assessment targeting students' views about experimental physics -- the Colorado Learning Attitudes about Science Survey for Experimental Physics (E-CLASS) -- across multiple, required lab courses at a single institution.  We find that, while pseudo-longitudinal averages showed increases in students' E-CLASS scores in each consecutive course, analysis of longitudinal data indicates that this increase was not driven by a cumulative impact of laboratory instruction.  Rather, the increase was driven by a selection effect in which students who persisted into higher-level lab courses already had more expert-like beliefs, attitudes, and expectations than their peers when they started the lower-level courses.  
\end{abstract}

\maketitle

\section{\label{sec:intro}Introduction}

Laboratory courses represent a unique and potentially important component of the undergraduate physics curriculum.  These courses can provide students with opportunities to engage authentically with the process of experimental physics and learn practical lab skills.  One implicit measure of the extent to which these courses are valued by the physics community is that the majority of undergraduate physics programs in the United States include at least two required lab courses, and many require more than two.  However, historically, little work has been done clearly demonstrating the effectiveness of these courses at achieving their learning goals \cite{singer2012dber,wieman2015labs, holmes2017labs}.  Recently, demonstrating student learning in undergraduate lab courses has been the focus of additional attention from the physics education research community \cite{holmes2015criticalThinking, zwickl2013adlab, devore2016amplifiers}.  Studies within this body of literature have, to date, focused on measuring the effectiveness of individual lab courses or comparing the effectiveness of several similar courses.

While assessments of individual lab courses are necessary and useful, particularly with respect to improving student learning within those particular courses, another important element of assessing the effectiveness of these courses is to take a longitudinal look at student learning as they advance through multiple courses in the undergraduate lab curriculum.  A longitudinal investigation of this type could provide a more holistic view of whether and how student learning is enhanced as they complete successive lab courses.  This kind of longitudinal study has not been conducted previously in part because there are significant theoretical and practical barriers to such a study.  For example, there are many possible learning goals for physics lab courses, including reinforcing students' understanding of particular physics concepts \cite{wieman2015labs, singer2006labs, millar2004labs}; teaching students practical lab skills such as troubleshooting, critical thinking, and measurement skills \cite{dounasfrazer2017troubleshooting, holmes2015criticalThinking, AAPT2015guidelines, zwickl2013adlab}; and fostering students' understanding of, and appreciation for, the nature and importance of experimental physics \cite{trumper2003labs, zwickl2013adlab, millar2004labs, singer2006labs}.  Longitudinal investigations of improvement in students' conceptual knowledge over multiple lab courses would be difficult given that the majority of lab courses at different levels (e.g., introductory to advanced) target a wide range of physics topics with only small overlap in the particular physics concepts used.  While there is potentially more overlap in the practical lab skills targeted by courses at different levels of the curriculum, little work has been done developing strategies for assessing development of these skill within a single course \cite{singer2012dber} let alone across multiple courses at multiple levels.  

Alternatively, the goal of fostering students' views and understanding of experimental physics and its place within the discipline is applicable to lab courses at all levels of the undergraduate curriculum.  Additionally, an established and research-based assessment designed to measure improvements with respect to this goal in both introductory and advanced lab courses already exists.  This assessment -- known as the Colorado Learning Attitudes about Science Survey for Experimental Physics (E-CLASS) -- is described in more detail in Sec.\ \ref{sec:eclass}.  The importance of students' beliefs, attitudes, and expectations as a learning goal of lecture courses has been discussed extensively both for science education generally \cite{hofer1997epistemology, elby2009epistemology} as well as for physics education specifically \cite{adams2006class, redish1998mpex}.  Students' epistemologies and expectations have also been shown to be linked to other measures of student success such as performance, interest, and persistence in physics \cite{perkins2005class, perkins2006class, perkins2004class}.  Helping students to cultivate expert-like views with respect to the nature and importance of experimental physics has also been called out as an important goal of undergraduate physics labs by both researchers \cite{trumper2003labs, zwickl2013adlab, feynman1998goals} and national organizations \cite{AAPT2015guidelines, olson2012excel}.  

The goal of this paper is to present an example exploration of longitudinal trends in students' views and expectations about the nature of experimental physics in their lab courses across the lab curriculum.  Specifically, we utilize students' responses to the E-CLASS assessment collected over a period of roughly four years in each of the three required lab courses at a research university.  The primary goal of this analysis is to determine whether the observed increase in average E-CLASS scores in each successive lab course is due to the impact of instruction or the result of a selection bias related to which students choose to pursue and persist in the physics program.  After describing the courses, student population, and institution (Sec.\ \ref{sec:methods}), we present a longitudinal analysis of these data, which looks at individual students' scores as they progress through multiple lab courses (Sec.\ \ref{sec:results}).  We conclude with a discussion of the limitations and implications of our findings for undergraduate lab education (Sec.\ \ref{sec:discussion}).

\section{\label{sec:methods}Methods}

In this section, we present the assessment, data sources, student demographics, and analysis methods used for this study.  

\subsection{\label{sec:eclass}The E-CLASS assessment}

The E-CLASS is a research-based and validated survey that probes students' views about the nature and importance of experimental physics.  In the E-CLASS, students are asked to rate their level of agreement -- from strongly agree to strongly disagree -- to 30 statements, such as, ``When doing an experiment, I just follow the instructions without thinking about their purpose.''  Students respond to each statement both from their own perspective when doing experiments in their laboratory course, and from the perspective of a hypothetical experimental physicist.  Numerical scores on each of the E-CLASS items were determined based on the established expertlike response for that item \cite{zwickl2014eclass}.  For each item, the responses `(dis)agree' and `strongly (dis)agree' were collapsed into a single `(dis)agree' category, and students were awarded $+1$ for favorable (i.e., consistent with experts), $+0$ for neutral, and $-1$ for unfavorable (i.e., inconsistent with experts).  A student's overall E-CLASS score is then given by the sum of their scores on each of the 30 items resulting in a possible range of scores of $[-30,30]$ \cite{wilcox2016eclass}.  

Over approximately the last four years, the E-CLASS has been administered at more than 75 different institutions primarily in the United States via a centralized, online administration system \cite{wilcox2016admin}.  This national data set includes both introductory and upper-division courses, and a variety of different institution types. We have previously presented the theoretical grounding, development, and validation of the E-CLASS \cite{zwickl2014eclass, wilcox2016eclass}.  We have also utilized this extensive data set to explore the role gender plays in performance on E-CLASS \cite{wilcox2016gender}. Additionally, we measured the impact of different types of lab activities on E-CLASS scores, where we found students in courses that included at least some open-ended activities outperformed students in courses with only guided labs \cite{wilcox2016structure}. Similarly, we have measured a significant improvement in E-CLASS scores in courses that use well-established transformed curricula compared to traditional labs at the introductory level, and the increase is significantly larger for women \cite{wilcox2016pedagogy}. Finally, we have shown that courses that focus more on developing lab skills outperform courses that focus more on reinforcing physics concepts, and again, the increase is significantly larger for women \cite{wilcox2016focus}.

\subsection{\label{sec:data}Data sources and context}

Data for this study were collected over eight consecutive semesters at the University of Colorado Boulder (CU).  The undergraduate physics and engineering physics curriculum at CU includes three required lab courses, one at each the freshman, sophomore, and junior years, along with one additional optional lab course at the senior level.  The senior-level, Advanced lab course at CU covers topics in optics and modern physics and is offered as an alternative path for students to fulfill the requirement that all CU physics majors engage in research in order to graduate.  Thus, students who take this course are almost exclusively physics students who have not had an undergraduate research experience either because they did not want one or because they were unable to secure one.  The clear selection bias introduced by the optional nature of this senior lab course presents a significant challenge with respect to interpreting longitudinal trends; thus, this course has been removed from the overall data set.  The following analysis will focus on data from CU's three required lab courses described below.  

\begin{table}
\caption{Breakdown of majors in each of CU's three required lab courses.  Here, physics includes both physics and engineering physics majors; engineering includes all engineering majors except engineering physics; other science includes both math majors and all science majors except physics and engineering; and non-science includes both majors who are declared non-science majors as well as students who are open-option/undeclared.}\label{tab:majors}
\begin{ruledtabular}
   \begin{tabular}{ l c c c c c }
     & & & Other & Non- & \\
	Course & Physics & Engineering & Science & Science & Total \\
     \hline
     1140 	& 11\% & 55\%	& 30\% 	& 3\% & 2306  \\
     2150 	& 71\% & 4\% 	& 23\% 	& 1\% & 188  \\
     3330	& 94\% & 0\%	& 5\% 	& 0\% & 221  \\
   \end{tabular}
\end{ruledtabular}
\end{table}  

The introductory lab course at CU (PHYS 1140) is a one credit course typically taken in the second semester of students' freshman year.  While all physics and engineering physics majors are required to take this course, the student population is primarily engineering majors and other science majors such as biology, chemistry, and astrophysics (see Table \ref{tab:majors}).  Students in PHYS 1140 complete six, two-week long experiments related to topics in introductory mechanics as well as electricity and magnetism.  The second lab course (PHYS 2150) is also a one credit course typically taken during students' sophomore year.  The student population is primarily physics and engineering physics majors with smaller number of general engineering and other science majors (see Table \ref{tab:majors}).  Students in PHYS 2150 choose 6 of 12 possible modern physics experiments to complete throughout the semester.  Both PHYS 1140 and 2150 are taught in a traditional guided lab format and students' grade is primarily based on submission of a formal lab report.  The final required lab course at CU (PHYS 3330) is a two credit course typically taken during the students' junior year.  The student population is almost exclusively physics and engineering physics majors (see Table \ref{tab:majors}).  Students in PHYS 3330 complete 10 guided lab activities and then the final 4-5 weeks of the semester are dedicated to a single open-ended project.  

The E-CLASS was administered online both pre- and postinstruction during all eight semesters of both PHYS 1140 and 3330.  Due to logistical issues outside the control of the research team, the E-CLASS was only administered both pre- and postinstruction during five of the eight semesters in PHYS 2150.  Instructors for the courses typically offered a small amount of participation credit to encourage students to complete the assessment.  Instructors never received students' raw responses; instead, they received a list of students who participated and a summary of their students' responses in aggregate.  Students' responses were matched, both pre- to postinstruction and between courses, first by student ID numbers and then by first and last name when ID matching failed.  

\begin{table}[b]
\caption{Number of students in each of the three partially longitudinal populations in our data set.  The 33 students for whom we have fully longitudinal are included in all applicable partially longitudinal populations, and thus appear multiple times in these counts.  }\label{tab:longN}
\begin{ruledtabular}
   \begin{tabular}{ l c c }
					& PHYS 1140		& PHYS 2150	 \\
	\hline
	PHYS 2150 	& 106				& -- \\
	PHYS 3330 	& 61 				& 63 \\
   \end{tabular}
\end{ruledtabular}
\end{table}  

Within our data set, there are both students who are fully longitudinal and students who are partially longitudinal (see Table \ref{tab:longN}).  By fully longitudinal, we mean that these students have matched pre- and postinstruction responses in each of the three lab courses.  Only 33 students in our data set are fully longitudinal.  Partially longitudinal students are students who have matched E-CLASS responses from at least two of the three lab courses (note that this means the fully longitudinal students are also included in the partially longitudinal population).  Students in the partially longitudinal population include students for whom we have pre- and postinstruction responses from: both PHYS 1140 and 2150 ($N=106$), both PHYS 2150 and 3330 ($N=63$), or both PHYS 1140 and 3330 ($N=61$).  Note that any student for whom we have both PHYS 1140 and 3330 data had to have taken and passed PHYS 2150.  Thus, regardless of whether we have matched E-CLASS data from them in PHYS 2150, these students also provide a fully longitudinal view of whether (and how) students E-CLASS scores change over the course of all three lab courses at CU.  In the analysis that follows we will examine each of these three longitudinal populations separately and compare them with the full pseudo-longitudinal population of each course.  Here, the pseudo-longitudinal population refers to the full population of students in each course; in other words, the pseudo-longitudinal analysis looks at changes in the overall averages of consecutive courses without attempting to match students between courses.

\section{\label{sec:results}Results}

\begin{table}[b]
\caption{Pseudo-longitudinal E-CLASS averages (max score 30) for all students who took each lab course during the data collection period.  Pre- to postinstruction shifts are statistically significant (Mann-Whitney U \cite{mann1947mwu} $p<0.05$) for both PHYS 1140 and 2150. }\label{tab:pseudo}
\begin{ruledtabular}
   \begin{tabular}{ l c c c }
			& PHYS 1140		& PHYS 2150	 & PHYS 3330\\
	\hline
	Pre 	& 18.1				& 19.5		& 20.1 \\
	Post 	& 15.8 			& 17.6 	& 20.2 \\
	N	    &  2306			& 188		& 221  \\
   \end{tabular}
\end{ruledtabular}
\end{table}  

\begin{table*}
\caption{Overall E-CLASS averages (max score 30) for students in each of the three longitudinal populations.  The longitudinal students (L) represent student for whom we have matched pre- and postinstruction responses in both courses in question.  The general students (G) represent all students for whom we have matched pre- and postinstruction responses for only the given course (minus the longitudinal students).  Differences between scores for longitudinal and general students in a given course are statistically significant (Mann-Whitney U $p<0.05$) for the lower level of the two course in all cases except one (denoted by asterisks), while the same difference for the higher level of the two courses is statistically insignificant in all cases.  }\label{tab:longTot}
\begin{ruledtabular}
   \begin{tabular}{ l c c c c c c c c c c c c c c c  }
     && \multicolumn{4}{c}{PHYS 1140 to 2150} &&\multicolumn{4}{c}{PHYS 1140 to 3330} && \multicolumn{4}{c}{PHYS 2150 to 3330}\\
	&& \multicolumn{2}{c}{1140} & \multicolumn{2}{c}{2150} && \multicolumn{2}{c}{1140} & \multicolumn{2}{c}{3330} && \multicolumn{2}{c}{2150} & \multicolumn{2}{c}{1140} \\
	Course && L & G & L & G && L & G & L & G && L & G & L & G \\
     \hline
     Pre	&& 20.4 & 18.0 & 19.6 & 19.4 && 21.0 & 18.0 & 20.6 & 20.0 && 20.6* & 18.9* & 20.3 & 20.1 \\
     Post 	&& 19.3 & 15.7 & 17.3 & 18.1  && 21.0 & 15.7 & 21.0 & 20.0 && 18.8 & 16.9 & 20.6 & 20.1 \\
     N	   	&& 106 & 2220 & 106 & 82  && 61 & 2245 & 61 & 160 &&  63 & 125 & 63 & 158 \\
   \end{tabular}
\end{ruledtabular}
\end{table*}  

If selection effects as students advance through the curriculum are not significant, then pseudo-longitudinal data would be sufficient to determine the impact of the lab curriculum on students' E-CLASS scores.  This is an attractive possibility as pseudo-longitudinal data are significantly easier and faster to collect.  Table \ref{tab:pseudo} reports the overall pre- and post instruction E-CLASS averages for the full population of students who took each lab course during the period of data collection.  There are several interesting features of these data, including the fact that all courses show a negative or neutral shift from pre- to postinstruction, with no positive shifts.  In contrast, Table \ref{tab:pseudo} also indicates a steady increase in students preinstruction averages in the higher level courses.  This motivates the question, what is the source of these preinstruction increases given that the overall shift in individual courses is negative or neutral?  To address this question, we examine scores of students for whom we have longitudinal data.  

\begin{figure}
\includegraphics[width=\linewidth]{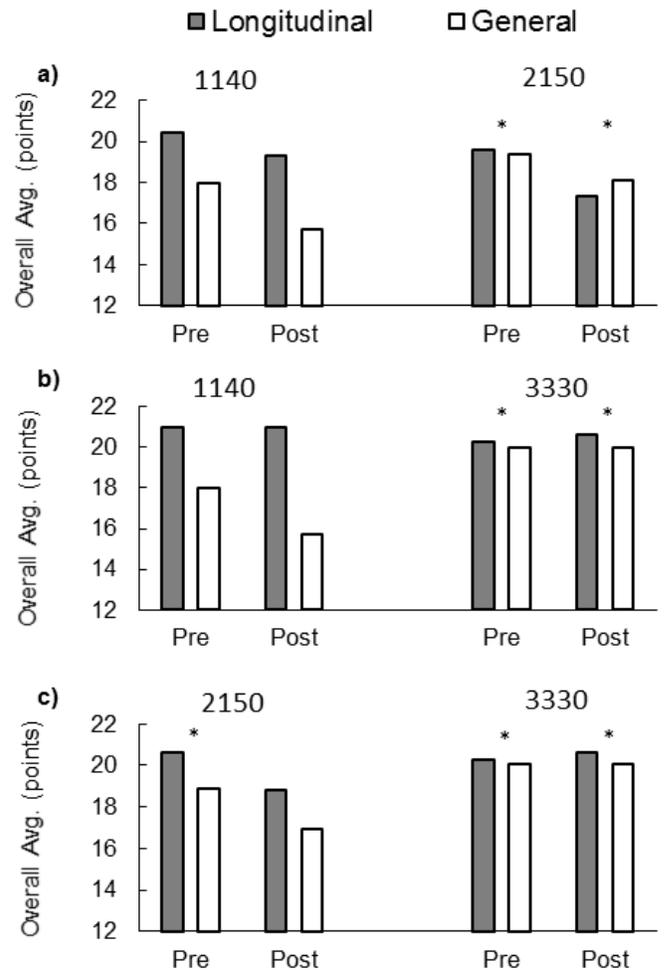}
\caption{Visual representation of the data included in Table \ref{tab:longTot} representing scores from: a) the PHYS1140 to PHYS 2150 comparison, b) the PHYS 1140 to PHYS 3330 comparison, and c) the PHYS 2150 to PHYS 3330 comparison.  Consistent with Table \ref{tab:longTot}, asterisks represent differences between the scores of the longitudinal and general population that are not statistically significant.  Note that the y-axis in all graphs has been truncated at 12 pts to facilitate visual comparisons of the differences in scores. }\label{fig:longTot}
\end{figure}

Table \ref{tab:longTot} presents the average E-CLASS scores for students in each of the longitudinal populations, and a visual representation of these data is given in Fig.\ \ref{fig:longTot}.   For all three populations, the students for whom we have longitudinal data start and end the lower level course with higher E-CLASS scores than the general population in that course.  This difference in scores is statistically significant (Mann-Whitney U \cite{mann1947mwu} $p<0.05$) in all cases except one -- the preinstruction score for the longitudinal students in PHYS 2150 is statistically the same as the preinstruction score for the general 2150 students.  Moreover, for all three populations, the longitudinal students start and end the higher level course with overall E-CLASS scores that are statistically indistinguishable (Mann-Whitney U $p>0.05$) from the scores of the general population of that course.  These trends suggest that the increase in students' scores in the higher level lab courses observed in the pseudo-longitudinal data (see Table \ref{tab:pseudo}) is strongly, if not entirely, driven by a selection effect.  Students that persist into higher level courses, started out more expert-like than their peers who did not persist, and their views stayed roughly constant throughout the curriculum.  

Because only the physics and engineering physics majors are required to take courses beyond PHYS 1140, these students are over-sampled in the longitudinal PHYS 1140 populations relative to the general 1140 population.  Thus, there are two potential selection effects at work, one related to which students select into the physics and engineering physics major and one to which students persist between courses independent of major.  Parsing out the extent to which each of these selection effects comes into play could include an analysis of longitudinal students who are physics majors relative to those who are not.  Our data set is not large enough to make strong statistical claims about the longitudinal physics and non-physics populations separately.  However, a preliminary look at the general trends in our sample showed that, other than the physics students having slightly higher averages overall, the longitudinal trends do not show signs of being meaningfully different between these two populations.  If robust, this finding would suggest that a selection effect relating to which students persist between courses was significant independent of major.

\section{\label{sec:discussion}Summary and Conclusions}

Using a research-based assessment -- know as E-CLASS -- we investigated longitudinal trends in students' beliefs about the nature and importance of experimental physics as they progress through the required laboratory curriculum at a large, very high research institution.  We collected pre- and postinstruction E-CLASS data from eight semesters of three consecutive lab courses that are required for all physics and engineering physics majors.  Using these data, we demonstrated that, while pseudo-longitudinal averages showed increases in students' E-CLASS scores in each consecutive course, analysis of longitudinal data indicates that this increase was not driven by a cumulative impact of laboratory instruction, but rather a selection effect in which students who persisted into higher level lab courses had started their lower level courses with more expert-like beliefs, attitudes, and expectations (as measured by E-CLASS) than their peers.  This finding is consistent with prior work investigating longitudinal trends a related assessment targeting students attitudes and beliefs in lecture physics courses, the Colorado Learning Attitudes about Science Survey (CLASS) \cite{adams2006class, gire2009classLong}.

One important implication of this finding is to underscore the importance of truly longitudinal data when investigating the impact of instruction across multiple courses.  While previous studies have found matching trends between longitudinal and pseudo-longitudinal data \cite{slaughter2011long}, our findings suggest that this is not always the case and should not be assumed.  The work described here has several important limitations.  This investigation was conducted at a single institution with a fairly traditional lab curriculum.  Our findings may not generalize to physics programs that have a significantly different student population or lab curriculum.  For example, our prior work investigating E-CLASS scores at multiple institutions suggests that programs that include a significant number of open-ended activities throughout the curriculum, or utilize research-based instructional approaches may show different trends \cite{wilcox2016structure, wilcox2016pedagogy}.  However, our findings do suggest that for programs with a largely traditional lab curriculum, participation in these courses is not encouraging more expert-like views, but rather selecting in favor of students whose views are already consistent with those of experts.

\begin{acknowledgments}
This work was funded by the NSF-IUSE Grant No. DUE-1432204 and NSF Grant No. PHY-1125844.  Particular thanks to the members of PER@C for all their help and feedback.  
\end{acknowledgments}

\bibliography{master-refs-ECLASS-10_16}

\end{document}